\documentclass[aps,pra,twocolumn,a4paper,english]{revtex4}

\usepackage{enumerate}
\usepackage{amsmath}
\usepackage{graphicx}
\usepackage{array}
\usepackage{amssymb}
\usepackage{bm}
\usepackage{makecell}
\usepackage{multirow}
\usepackage{color}

\usepackage{algorithmic}
\usepackage{algorithm}
\usepackage{float}

\makeatletter
\newcommand{\removelatexerror}{\let\@latex@error\@gobble}
\makeatother

\begin{document}
\title{Efficient Depth Selection for the Implementation of Noisy Quantum Approximate Optimization Algorithm}

\author{Yu Pan$^{1,2}$}\email[]{ypan@zju.edu.cn}
\author{Yifan Tong$^{2}$}
\author{Shibei Xue$^{3,4}$}
\author{Guofeng Zhang$^5$}

\affiliation{$^1$State Key Laboratory of Industrial Control Technology, Zhejiang University, Hangzhou, 310027, P. R. China}
\affiliation{$^2$Institute of Cyber-Systems and Control, College of Control Science and Engineering, Zhejiang University, Hangzhou, 310027, P. R. China}
\affiliation{$^3$Department of Automation, Shanghai Jiao Tong University, Shanghai 200240, P. R. China}
\affiliation{$^4$Key Laboratory of System Control and Information Processing, Ministry of Education of China, Shanghai 200240, P. R. China}
\affiliation{$^5$Department of Applied Mathematics, The Hong Kong Polytechnic University, Hung Hom, Kowloon, Hong Kong Special Administrative Region of China, P. R. China}

% USE FOR BOTH
\begin{abstract}
Noise on near-term quantum devices will inevitably limit the performance of Quantum Approximate Optimization Algorithm (QAOA). One significant consequence is that the performance of QAOA may fail to monotonically improve with depth. In particular, optimal depth can be found at a certain point where the noise effects just outweigh the benefits brought by increasing the depth. In this work, we propose to use the model selection algorithm to identify the optimal depth with a few iterations of regularization parameters. Numerical experiments show that the algorithm can efficiently locate the optimal depth under relaxation and dephasing noises.
\end{abstract}

\maketitle

\section{Introduction}
In the absence of effective and scalable error correction techniques, the current research on near-term Noisy Intermediate Scale Quantum (NISQ) \cite{Preskill18} computing has been focused on Quantum Approximate Optimization Algorithm (QAOA) which is expected to achieve quantum advantage for practical application \cite{farhi2014quantum,farhi2015quantum,farhi2019quantum}. QAOA is a hybrid quantum-classical algorithm for solving a special class of problems such as combinatorial optimization. An approximate solution for such problems would be acceptable as they have been proven to be NP-hard \cite{ZW20,borle2021quantum}.

QAOA consists of a quantum control protocol whose variational control parameters are updated according to the measurement result of each iteration. During each iteration, two noncommuting Hamiltonian controls are alternatively applied on the quantum system, for which the control parameters can be understood as angles of rotations or control durations. Each Hamiltonian control is then implemented as a circuit made up of quantum gates. Due to the high noise level and gate imperfection in the NISQ devices, the performance of QAOA critically relies on the number of variational control parameters, which is defined as the control depth in this paper. Nevertheless, the current literature often assumes a noise-free model \cite{farhi2014quantum,JRW17,hastings2018,WHJR18,ZW20,larkin2020,BARE20,LHH21}. In particular, the performance of QAOA will monotonically increase with the control depth \cite{farhi2014quantum} if noise is not taken into consideration. However, a large depth is not realistic for NISQ devices due to the prominent noise effect and limited hardware capacity. Therefore, a very small depth, which is usually determined by empirical rules, has to be chosen in these works to validate the noise-free assumption.

QAOA will inevitably confront the need to increase the control depth as the problem to solve becomes more and more complex. In that case, noise effect has to be accounted for and mitigated to improve the overall performance. Although circuit design, gate optimization and increasing the freedom of classical control \cite{MAS20,alam2020circuit,PRXQuantum1020304,herrman2021multiangle} have been proposed to reduce the control depth and counteract the noise effect, it is still not clear that how to select an optimal control depth under the influence of a specific noise. Moreover, since the algorithm is actually implemented on quantum hardware, the noise is often unknown and the control parameters may have to be optimized via a data-driven way. It should be noted that one does not need to know the underlying model of the quantum system for the execution of QAOA. Hence, QAOA can be taken as a type of iterative learning control \cite{daoyi20} which is naturally adaptive to noise. More explicitly, during each iteration, the control parameters are updated based on the measurement results (e.g., \cite{farhi2014quantum,guerreschi2017,verdon2019,Sweke2020,Daniel20}) to maximize the optimization accuracy no matter what noise the quantum system has been exposed to.

Although extensive research efforts have been devoted to developing algorithms for optimizing the control parameters, the automatic optimization of control depth under quantum noise has not been studied so far. It is commonly known that the performance of a learning or optimization model mainly lies on the selection of hyperparameters, while control depth is the most important hyperparameter for QAOA \cite{APC20,CRAB21,HOH21}. In this paper, we propose an automatic algorithm for optimizing the control depth of noisy QAOA. To be more precise, we adopt the $l_1$ regularization technique to reduce the preselected large control depth to an optimal value during the iteration. This approach has been widely used to enforce sparse solutions \cite{JWHT13} in statistical learning. The sparsity, or complexity of the $l_1$-regularized model, has to be controlled by a proper criterion. For example, Akaike’s Information Criteria (AIC) and Bayesian Information Criteria (BIC) \cite{BA02} can be used to find the balance between a good optimization accuracy and complexity of the control depth. For noisy QAOA, instead of designing an empirical criterion for model selection, the optimal depth is naturally constrained. Due to the inevitable noise in NISQ devices, the computation errors will accumulate over time and optimization accuracy will decrease with the depth after reaching a peak. Particularly, it has been found by numerical simulations and experiments in \cite{Marshall_2020,MAS20,HSM21} that the performance of QAOA may not monotonically increase with depth under noise. As a result, the optimal depth can be simply defined as the one achieving the highest accuracy. The optimal depth with the highest accuracy is then found by solving an $l_1$-regularized optimization problem with a fast algorithm, which is much more efficient than exhaustive search.

This paper is organized as follows. A brief introduction to noisy QAOA and its $l_1$ regularized variant is given in Section \ref{Sec:NQAOA}. The algorithm for depth optimization is given in Section \ref{Sec:PG}. In Section \ref{Sec:NR}, the Max-Cut example, metrics and numerical results are presented. Section \ref{Sec:C} concludes this paper.

\section{Standard and Noisy QAOA}\label{Sec:NQAOA}
The noise-free QAOA is executed as a sequence of unitary control, which is generated by applying a piece-wise Hamiltonian as
\begin{equation}
H:=H_o\stackrel{\gamma_1}{\longrightarrow}H_c\stackrel{\beta_1}{\longrightarrow}\ \cdot\cdot\cdot\ H_o\stackrel{\gamma_p}{\longrightarrow}H_c\stackrel{\beta_p}{\longrightarrow}.\label{Hdef}
\end{equation}
Here $H_o$ is the problem-based Hamiltonian whose ground states encode the solution to the optimization problem, and $H_c=\sum_{n=1}^N\sigma_{x}^{(n)}$ is the noncommuting control Hamiltonian. The variational parameters $\beta=(\beta_1,\cdot\cdot\cdot,\beta_p),\gamma=(\gamma_1,\cdot\cdot\cdot,\gamma_p)$ can be understood as control durations or angles of rotations. $2p$ is the initial depth of control sequence. We have used the following notations
\begin{equation}
\sigma_z=\left(
\begin{array}{cc}
 1& 0 \\
 0& -1
\end{array}
\right),\ \sigma_x=\left(
\begin{array}{cc}
 0& 1 \\
 1& 0
\end{array}
\right),\ \sigma_-=\left(
\begin{array}{cc}
 0& 0 \\
 1& 0
\end{array}
\right).
\end{equation}
The computational basis states of a single qubit are written as $|0\rangle=(0\quad 1)^T$ and $|1\rangle=(1\quad 0)^T$. During each iteration, the final state is generated by the Hamiltonian control as
\begin{equation}
|\psi(\beta,\gamma)\rangle=U(H_c,\beta_p)U(H_o,\gamma_p) \cdot\cdot\cdot U(H_c,\beta_1)U(H_o,\gamma_1)|s\rangle,\nonumber\\
\label{cp}
\end{equation}
where we have defined the unitary evolutions as $U(H_c,\beta_j)=e^{-\mbox{i}H_c\beta_j}$ and $U(H_o,\gamma_j)=e^{-\mbox{i}H_o\gamma_j}$. The initial state is $|s\rangle=|+\rangle_1\cdot\cdot\cdot|+\rangle_N$, with $|+\rangle=1/\sqrt{2}(|0\rangle+|1\rangle)$. The standard QAOA attempts to find the approximate solution to the following problem
\begin{equation}
\min_{x=(\beta,\gamma)}f(x)=\min_{x}\langle\psi(x)|H_o|\psi(x)\rangle,\label{pmin}
\end{equation}
where $f(x)$ is the expectation of $H_o$ taken on the final state $|\psi(x)\rangle$. Since $H_o$ can be decomposed as a combination of quantum projective measurements, $f(x)$ can be estimated by repeating the quantum control procedures for many times and taking the average of measurement results. Clearly, the state preparation, evolution and measurement can be implemented on quantum hardware, while the control parameters are optimized using a classical algorithm.

As QAOA is expected to run on noisy quantum hardware, the practical optimization accuracy may not monotonically increase with the depth under the influence of quantum noise. Specifically, the state of an open quantum system is described by a density matrix instead of a pure state vector. In order to study the influence of the noise, we simulate the dynamical evolution of the open quantum system using quantum master equation. The general formulation of quantum master equation in the Lindblad form reads \cite{BP07}
\begin{equation}
\dot{\rho}=-\mbox{i}[H,\rho]+\sum_{n=1}^N\gamma_n(L_n\rho L_n^\dagger-\frac{1}{2}L_n^\dagger L_n\rho-\frac{1}{2}\rho L_n^\dagger L_n),\label{meq}
\end{equation}
with $\gamma_n$ being the coupling strength between the $n$-th qubit and its environment. We assume that each qubit is independently coupled to the environment via the coupling operator $L_n$. $H$ is the piece-wise control Hamiltonian defined in (\ref{Hdef}) that alternates between $H_o$ and $H_c$. It is clear from (\ref{meq}) that the decay of a quantum state will accumulate as the total control time increases.

The initial state is then written as $\rho_0=|s\rangle\langle s|$. The state generated at the end of each iteration is denoted as $\rho(x)$. It should be noted that in this case $H_o$ is measured against $\rho(x)$ instead of a pure state, and the measurement result is calculated by $\mbox{tr}(H_o\rho(x))$. The regularized model for noisy QAOA is thus given by
\begin{eqnarray}
\min_{x}\mbox{tr}(H_o\rho(x))+\lambda||x||_1,\label{noiseobj}\\
||x||_1=\sum_{j=1}^p(|\beta_j|+|\gamma_j|)\nonumber,
\end{eqnarray}
where $||\cdot||_1$ refers to the $l_1$ norm, and $\lambda>0$ is a regularization parameter.

\section{Algorithm}\label{Sec:PG}
\subsection{Proximal Gradient Descent}
The unregularized version of (\ref{noiseobj}) can be solved simply by calculating its numerical gradient as \cite{guerreschi2017,WHJR18,Sweke2020,Daniel20}
\begin{equation}
[\nabla \mbox{tr}(H_o\rho(x))]_i\approx\frac{\mbox{tr}(H_o\rho(x_i-\epsilon))-\mbox{tr}(H_o\rho(x_i+\epsilon))}{2\epsilon}\label{fd}
\end{equation}
with a small $\epsilon>0$. Here $[\nabla \mbox{tr}(H_o\rho(x))]_i$ denotes the $i$-th element of the gradient vector. The expectation $\mbox{tr}(H_o\rho(x_i\pm\epsilon))$ is obtained by perturbing the control parameter $x_i$ by a small amount of $\epsilon$ and measuring $H_o$ with respect to the generated states. The standard gradient update is then given by
\begin{equation}
x_{k+1}=x_k-\eta\nabla \mbox{tr}(H_o\rho(x_k)),
\end{equation}
where $\eta$ is the step size, and $k$ is the current iteration step. However, since the regularization term in (\ref{noiseobj}) is not differentiable, the standard gradient descent is not directly applicable to this problem.

\begin{figure}[!t]

    		\begin{algorithm}[H]
			\caption{Proximal Gradient Descent}
			\label{PG:implementation}
			\begin{algorithmic}[1]
				\FOR{$k = 1, \cdots, K$}
				\STATE $\quad z_k=x_k-\eta\nabla \mbox{tr}(H_o\rho(x_k));$
                \STATE $\quad x_{k+1}=S_{\lambda\eta}(z_k);\quad \backslash\backslash$ Soft thresholding
				\ENDFOR
                \RETURN $x_{K+1}$;
			\end{algorithmic}	
		\end{algorithm}

\end{figure}

Proximal Gradient (PG) descent \cite{AM09} provides a solution to the regularized problem (\ref{noiseobj}) by minimizing the quadratic approximation to $\mbox{tr}(H_o\rho(x_k))$ around $x_k$ as
\begin{eqnarray}
&&x_{k+1}=\nonumber\\
&&\mathop{\arg\min}_{y}\mbox{tr}(H_o\rho(x_k))+\nabla \mbox{tr}(H_o\rho(x_k))^T(y-x_k)\nonumber\\
&&+\frac{1}{2\eta}||y-x_k||_2^2+||y||_1,\label{pgmin}
\end{eqnarray}
in which the quadratic gradient $\nabla^2f(\cdot)$ is approximated by $I/\eta$ while leaving the regularization term $||y||_1$ alone. The solution to (\ref{pgmin}) can be obtained using PG descent detailed in Algorithm \ref{PG:implementation}, which introduces an additional soft-thresholding operation after the standard gradient descent.

The soft-thresholding operation is defined by
\begin{equation}
S_{\lambda \eta}({[z_k]}_i)=\left\{
\begin{array}{rcl}
&{[z_k]}_i-\lambda\eta,           & {[z_k]}_i>\lambda\eta \\
&0,    &     -\lambda\eta\leq{[z_k]}_i\leq\lambda\eta\\
&{[z_k]}_i+\lambda\eta,          & {[z_k]}_i<-\lambda\eta
\end{array} \right.
\end{equation}
The values of the control parameters are shifted towards zero by an amount of $\lambda\eta$ after soft thresholding. Moreover, once the condition $|{[z_k]}_i|\leq\lambda\eta$ is satisfied, the $i$-th control parameter will be penalized to exactly zero, which in turn reduces the control depth by removing the corresponding control action from the control sequence.

\subsection{Choice of $\lambda$}
The choice of regularization parameter $\lambda$ is very important for determining the depth. Generally speaking, a large $\lambda$ leads to a large threshold value $\lambda\eta$, and thus the control parameters are more likely to be penalized to zero during the iteration. On the contrary, a small $\lambda$ will keep most of the parameters in the model. Therefore, a set of values have to be tested in order to determine an appropriate regularization parameter that yields the optimal depth. To do this, for each $\lambda$, an optimized accuracy is obtained using the regularized QAOA with reduced depth. The optimized accuracies for different $\lambda$ are compared, and the optimal depth is the one that achieves the best accuracy.

The common practice to solve the $l_1$-regularized optimization problem is to start with a relatively large $\lambda$, which will shrink by a constant factor after each round of experiment. As $\lambda$ decreases, the selected depth in general will increase to improve the optimization accuracy. However, due to the noise effect, the optimization accuracy will reach a peak with a certain $\lambda$. Continuing decreasing $\lambda$ from that point is not necessary since it will introduce overwhelming noise, which makes this $\lambda$ optimal.

It should be noted that only a small number of values have to be tested for $\lambda$, which makes this algorithm very efficient. To be more specific, the regularization term and objective function $\mbox{tr}(H_o\rho(\cdot))$ should be comparable for the optimization to proceed. As a result, any effective $\lambda$ will be confined within a small range. After a few rounds of shrinkage, the regularization strength will soon become too small for depth selection.

\section{Numerical Results}\label{Sec:NR}

\subsection{Max-Cut and approximation ratio}
The performance of depth optimization is demonstrated on Max-Cut problem \cite{farhi2015quantum}. Consider an $N$-node non-directed and weighted graph $G=(V,E)$. Max-Cut is the partition of $V$ into two subsets $V_1$ and $V_2$, for which the sum of weights of edges between the nodes of two disjoint subsets is maximized. By assigning $1$ to the nodes of one subset and $-1$ to the nodes of the other subset, Max-Cut can be formulated as a binary optimization problem which is equivalent to the minimization of expectation of the corresponding Hamiltonian
\begin{equation}
H_o=\sum_{(i,j)\in E}\omega_{ij}\sigma_{z_i}\sigma_{z_j}.
\end{equation}
We define the approximation ratio as follows
\begin{equation}
r=1-\frac{\min_{x}\mbox{tr}(H_o\rho(x))-C_{\min}}{C_{\max}-C_{\min}}\in[0,1],\label{ar}
\end{equation}
with $C_{\min}$ being the theoretical minimum value and $C_{\max}$ being the theoretical maximum value. The approximation ratio $r$ is a measure of how close the final state is to the optimal solution, and a larger $r$ indicates a better solution.

\begin{figure}[!htp]
	\centering
	\includegraphics[width=9cm]{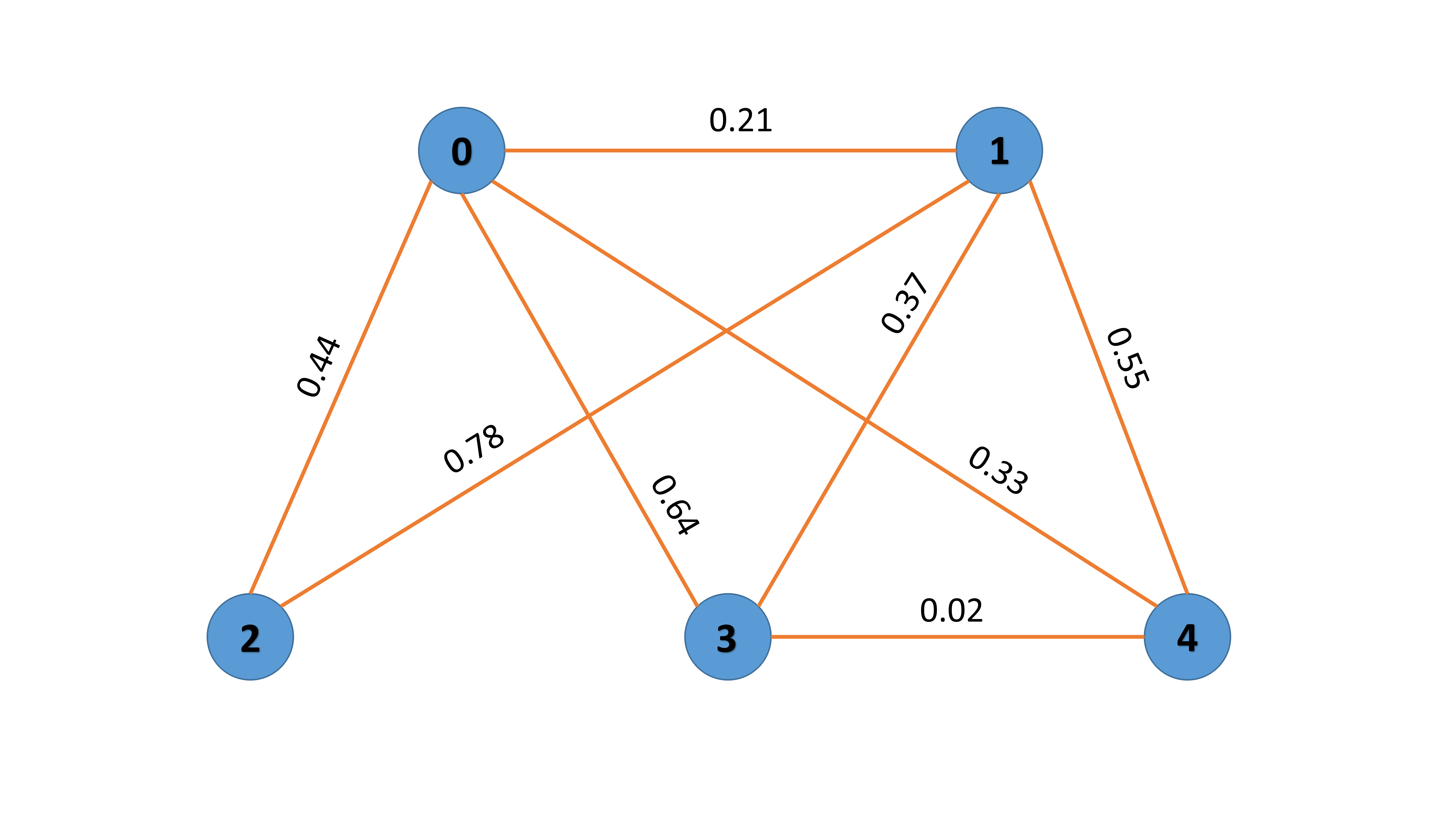}
	\caption{The randomly generated Max-Cut example used in the numerical experiment.}
	\label{fmaxcut}
\end{figure}

The Max-Cut problem to test in this paper is randomly generated as described in Fig.~\ref{fmaxcut}. The weighted graph is made up of 5 vertices and 8 edges. For simplicity, we set $\gamma=\gamma_1=...=\gamma_N$ in the simulation definition.

\begin{figure}[!htp]
	\centering
	\includegraphics[width=9cm]{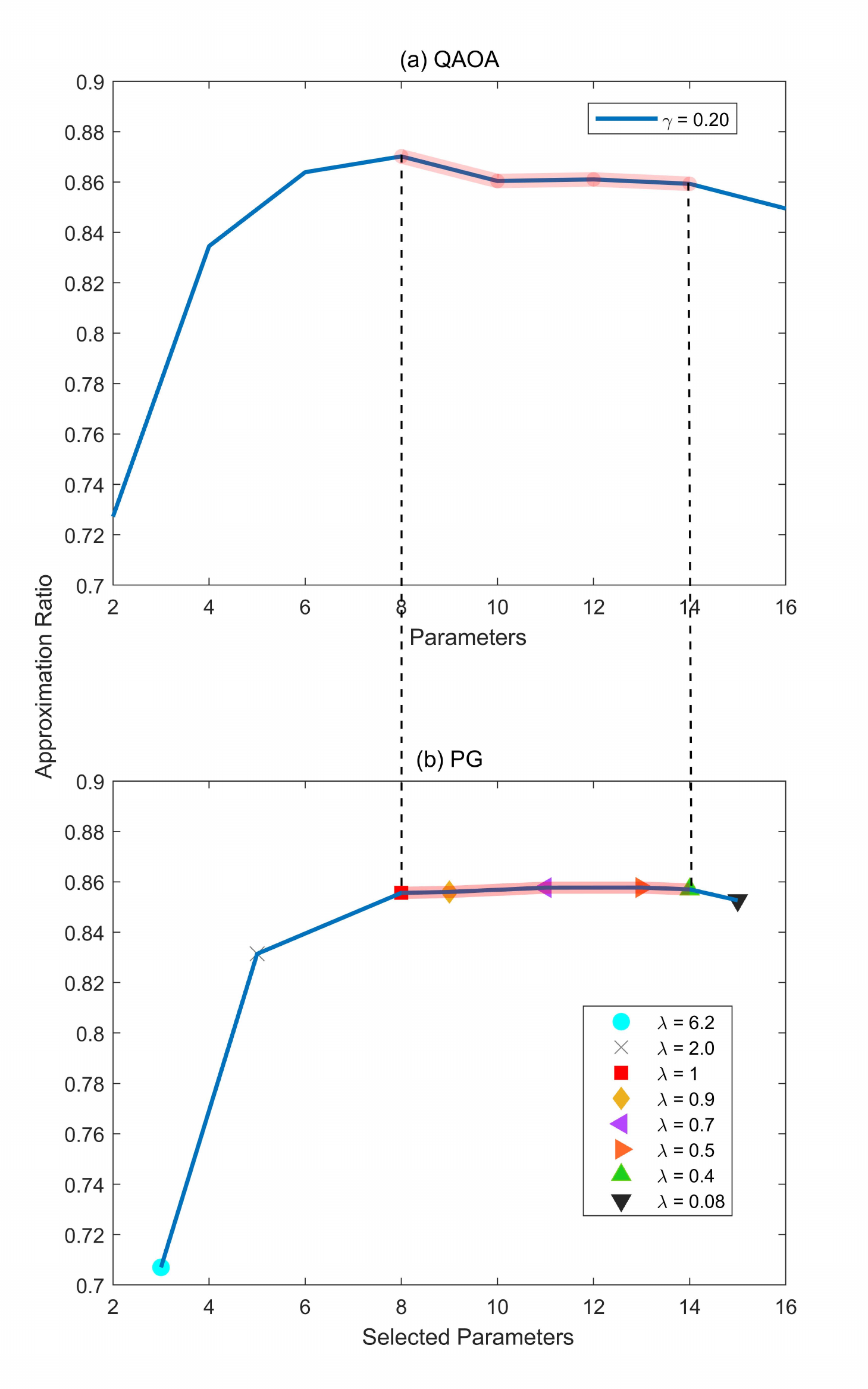}
	\caption{The performance comparison between (a) noisy QAOA and (b) PG algorithm under the influence of relaxation with $\gamma=0.2$. Each algorithm has been iterated for 300 times to obtain one approximation ratio. The approximation ratios achieved by noisy QAOA for different number of control parameters are depicted in (a), with an unconspicuous peak at $p=4$ followed by a slight downtrend till $p=7$. The PG algorithms all start with the same initial depth but different $\lambda$. (b) shows the finally selected number of parameters after optimization. The ideal number of control parameters lies in the region $p=[4,7]$, which is consistent with (a).}
	\label{fig.1}
\end{figure}

\subsection{Depth Selection with Relaxation Process}
The relaxation process is simulated by letting
\begin{equation}
L_1=...=L_n=\sigma_-.
\end{equation}
That is, the effect of noise for each qubit is to relax the system from the excited state to the ground state, which constitutes a major source of quantum decoherence.

The initial value of each control parameter is set as 0.1. The learning rate is 0.008. In addition, $H_o$ and $H_c$ are multiplied by a factor of 6 to accelerate the simulation of dynamics. As shown in Fig.~\ref{fig.1}, increasing the control depth for noisy QAOA does not always lead to performance improvement. On one hand, the noise effect accumulates with the growth of $p$, as the the total control duration increases with the depth when the initial value of each control parameter is fixed. On the other hand, it is increasingly hard to optimize the model with more parameters, which in general will slow down the pace of improvement once the approximation ratio has reached a certain level. At some point, the negative effects will outweigh the benefits of increasing the depth. In Fig.~\ref{fig.1}(a), the peak value of the approximation ratio is 0.8702, which has been achieved with 8 control parameters, corresponding to $p=4$. In addition, the curve shows a plateau after reaching the peak. Due to the relatively low noise level, it is possible that the increased noise brought by a few additional control steps can be counteracted with more freedoms in optimization.

The results of PG algorithm applied on the regularized model (\ref{noiseobj}) are shown in Fig.~\ref{fig.1}(b). The initial depth is $p=8$, which corresponds to 16 control parameters. The finally selected numbers of parameters are obtained with different values of $\lambda$. It can be seen that the algorithm can accurately identify the same plateau as Fig.~\ref{fig.1}(a) that indicates the ideal number of control parameters. In particular, the leftmost point on the plateau can be chosen as the solution, since it corresponds to the smallest model complexity. In practice, when such a plateau exists, we can always stop the experiment early if the difference between the last approximation ratio and the current value is less than a given threshold.

\begin{figure}[!htp]
	\centering
	\includegraphics[width=9cm]{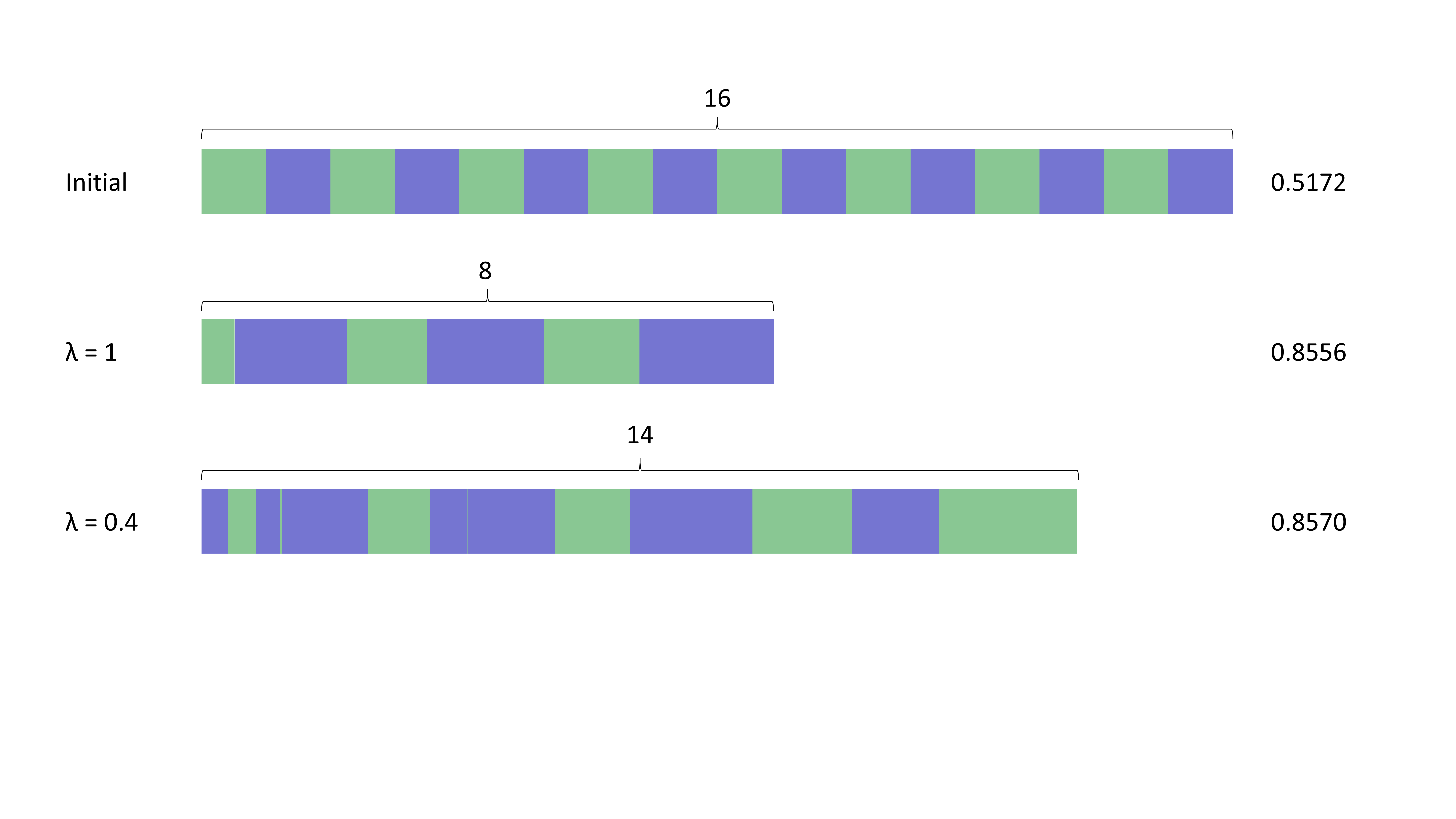}
	\caption{The initial control sequence, the finally selected control sequences with $\lambda=1$ and $\lambda=0.4$ are drawn. The numbers of control parameters for each sequence are 16, 8 and 14, respectively. The achieved approximation ratio is marked on the right. Different types of control operations are represented by different colors. Adjacent control operations of the same type are combined into a single operation.}
	\label{fpara}
\end{figure}

\begin{figure}[!htp]
	\centering
	\includegraphics[width=9cm]{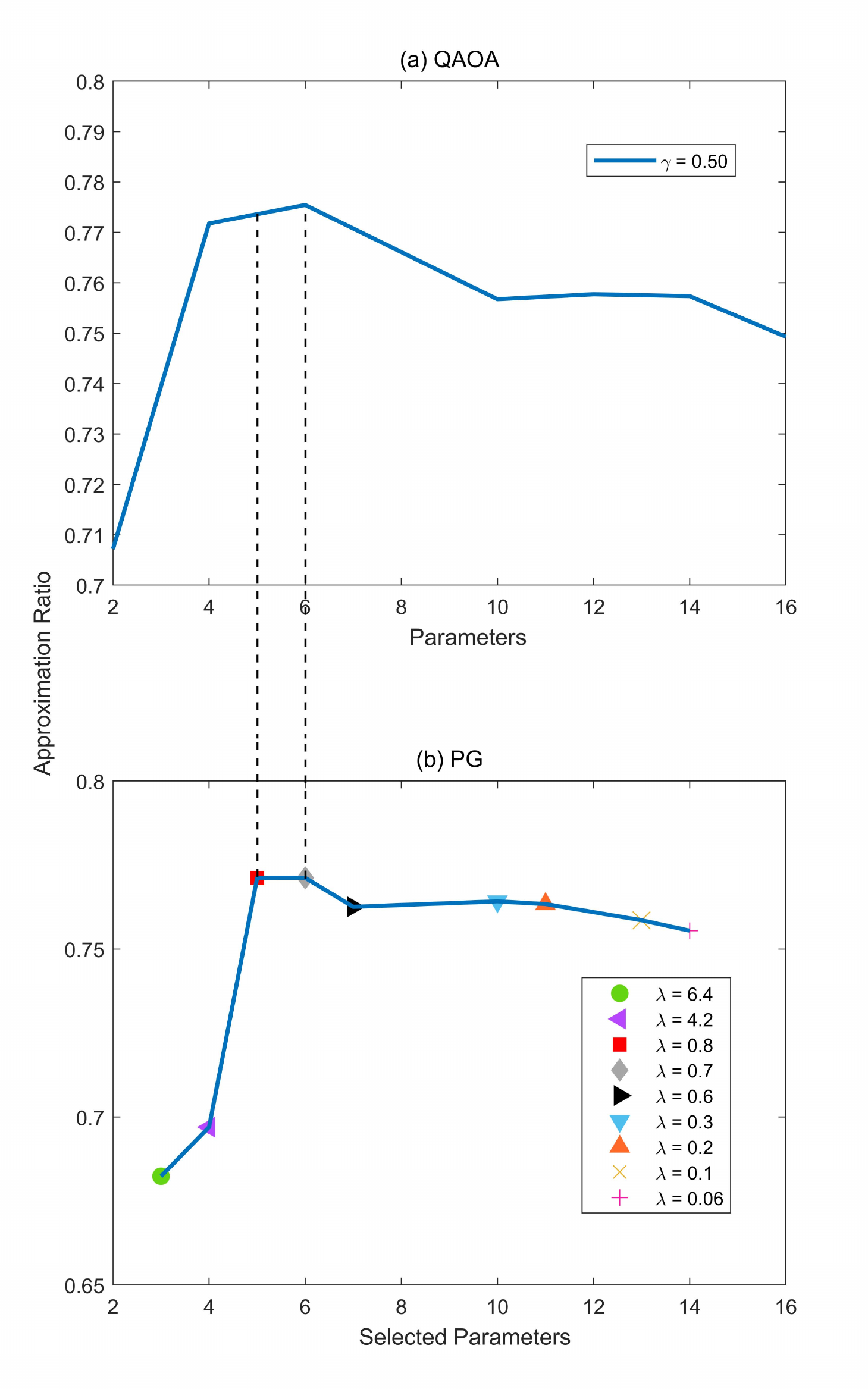}
	\caption{The performance comparison between (a) noisy QAOA and (b) PG algorithm under the influence of relaxation with $\gamma=0.5$. The approximation ratios achieved by 300 iterations are depicted in (a), with respect to different number of control parameters. (b) shows the finally selected number of parameters using the PG algorithms and 300 iterations. The approximation ratios are obtained with different $\lambda$, while the initial depths are the same. The selected number of control parameters is consistent with the best control depth of plain QAOA.}
	\label{fighn}
\end{figure}

The reduced depths and durations for $\lambda=1$ and $\lambda=0.4$ are visualized in Fig.~\ref{fpara}. It can be seen that although the approximation ratios are similar for the two cases, the control sequences are significantly different. In particular, the control sequence with 8 selected parameters is approximately $2/3$ the length of the control sequence with 14 selected parameters. It should be noted that if one control parameter is penalized to zero, then the corresponding control operation can be removed from the sequence. The control operations before and after the removed operation are of the same type, which means that they can be combined into one operation. Therefore, 8 selected parameters yield only 6 control operations, which is a significant reduction in depth when compared to the initial control sequence that involves 16 control operations.

\begin{figure}[!htp]
	\centering
	\includegraphics[width=9cm]{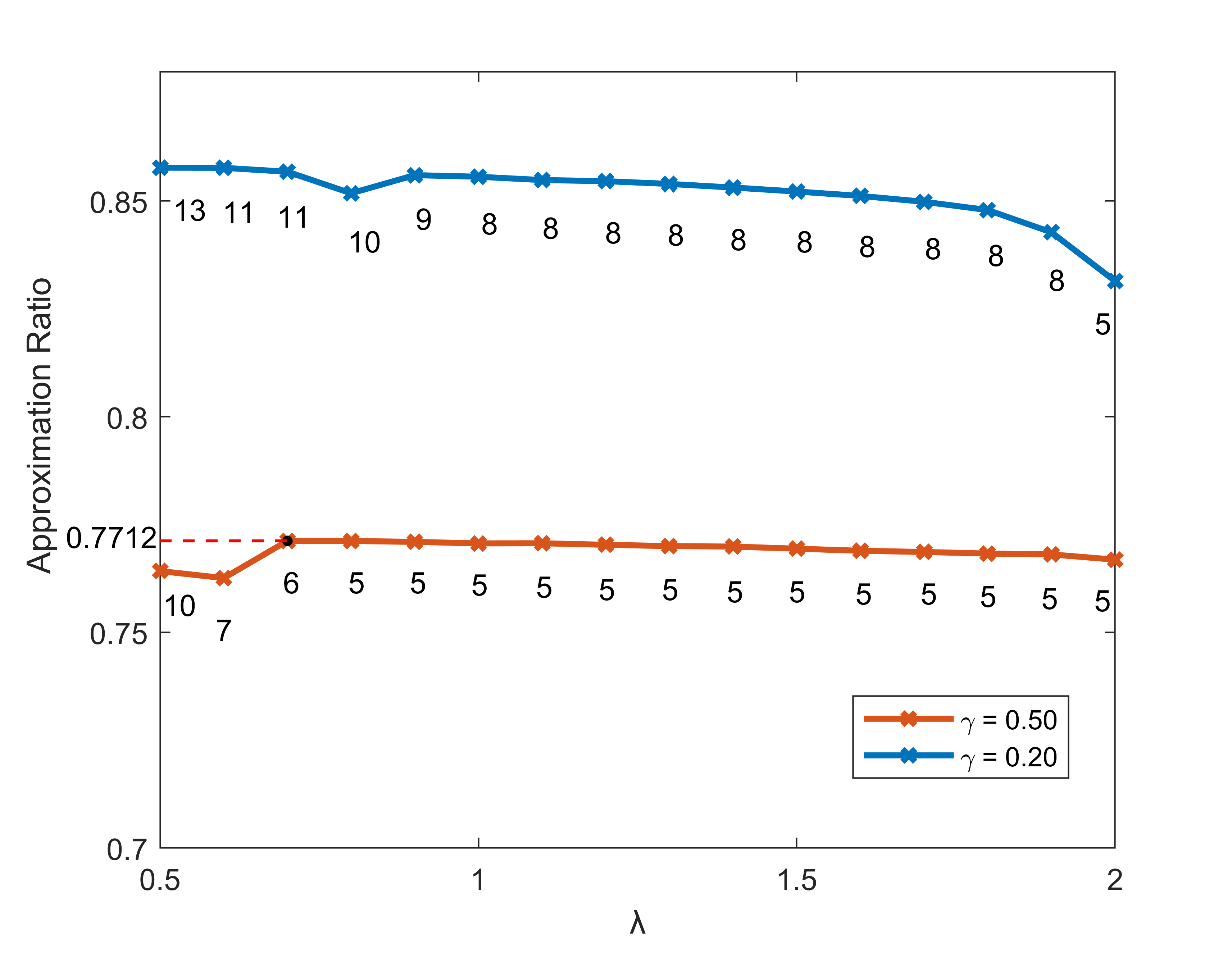}
	\caption{The orange curve of the finally selected number of parameters is obtained under the high noise level $\gamma=0.5$ by varying the value of $\lambda$ from 0.5 to 2, while the blue line is obtained under the low noise level $\gamma=0.2$.}
	\label{fig.2}
\end{figure}

As the noise level is increased, it becomes increasingly difficult to counteract the noise effect brought by the additional control steps. Therefore, once the peak is reached, the approximation ratio is expected to descend faster than the low-noise case. As shown in Fig.~{\ref{fighn}}(a), an obvious peak can be found when executing the QAOA with different number of control parameters. In particular, the optimal number of control parameters is either 5 or 6, which can be precisely identified using the PG algorithm.

As an intelligent algorithm, the PG algorithm is capable of efficiently determining the best depth with a small set of regularization parameters. That is, the selection result is robust to small variations in $\lambda$, such that a sparsely distributed set of regularization parameters are sufficient for finding the near-optimal solution. As shown in Fig.~{\ref{fig.2}}, the selection result is the same for any $\lambda$ that is confined in the region $\lambda\in(1,2)$. As a result, as long as we have tried one $\lambda$ in $(1,2)$, the near-optimal number of control parameters can always be found. For example, if the initial value of $\lambda$ is chosen as 6 which will be shrinked by a constant factor of 0.6 after each experiment, i.e., $\lambda=6\rightarrow 3.6\rightarrow 2.16\rightarrow 1.296 \rightarrow\cdot\cdot\cdot$, the optimal depth can be found with around 5 experiments.

\begin{figure}[!htp]
	\centering
	\includegraphics[width=9cm]{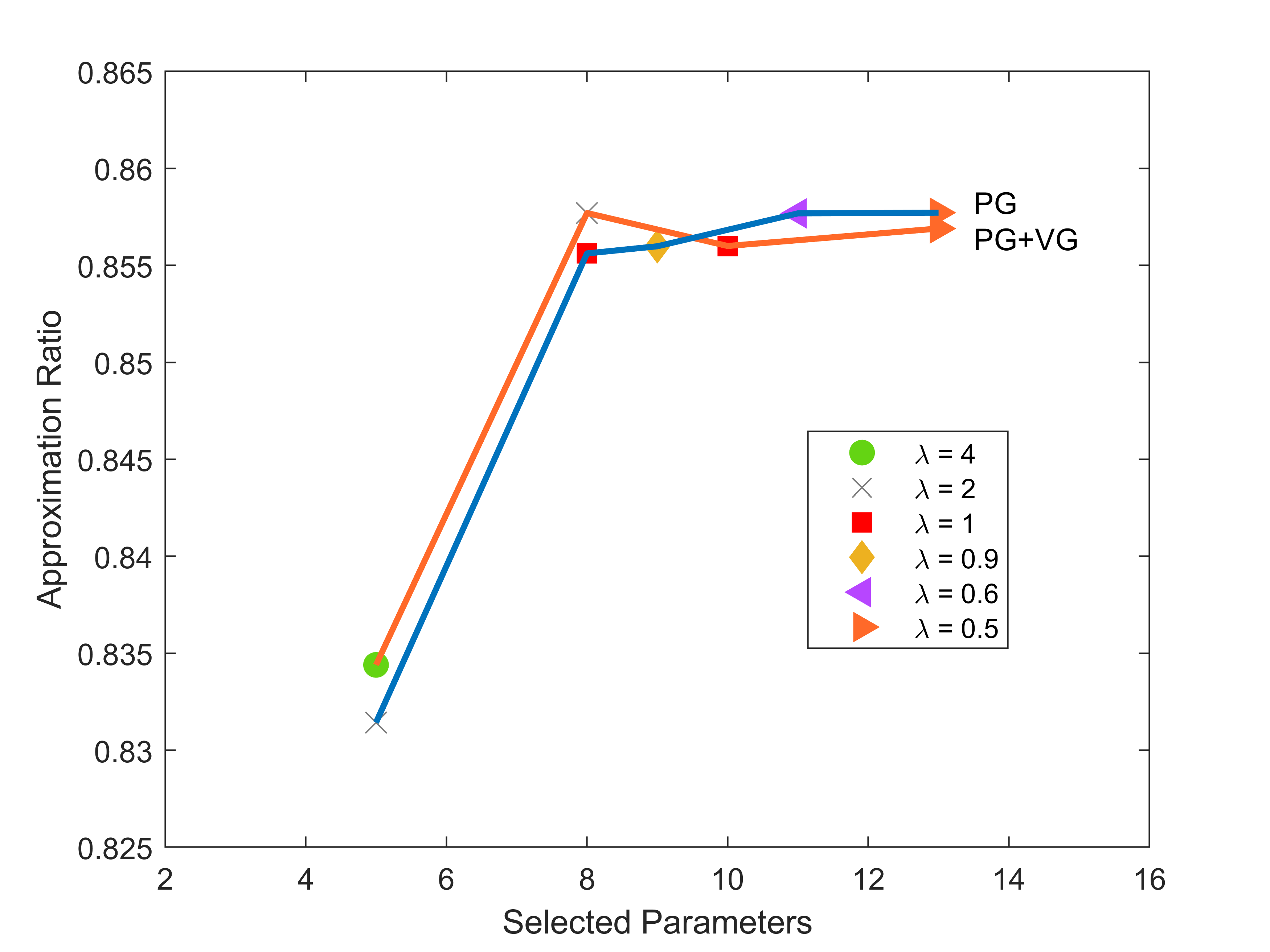}
	\caption{The noise level is $\gamma=0.2$. The orange curve shows the selected numbers of control parameters using 200 iterations of PG descent followed by 100 iterations of vanilla gradient descent. The blue curve shows the selected numbers of control parameters using 300 iterations of PG descent.}
	\label{fpvp}
\end{figure}

A more precise way to implement the depth selection method is by using the PG algorithm for the initial iterations, and then fixing the control depth for further optimization with vanilla gradient descent. By this way, the ultimate performance of QAOA with the selected depth can be precisely estimated, since the final accuracy is obtained without the regularization term. Here we test this implementation by using PG descent for the first 200 iterations and then employing vanilla gradient descent for the next 100 iterations. As shown in Fig.~\ref{fpvp}, an obvious peak appears on the curve of selected parameters which is in sharp contrast with the results obtained by 300 iterations of PG descent, and the peak exactly matches the best depth in Fig.~\ref{fig.1}(a). That is, the updated selection method could accurately identify the best depth instead of a plateau.

\begin{figure}[!htp]
	\centering
	\includegraphics[width=9cm]{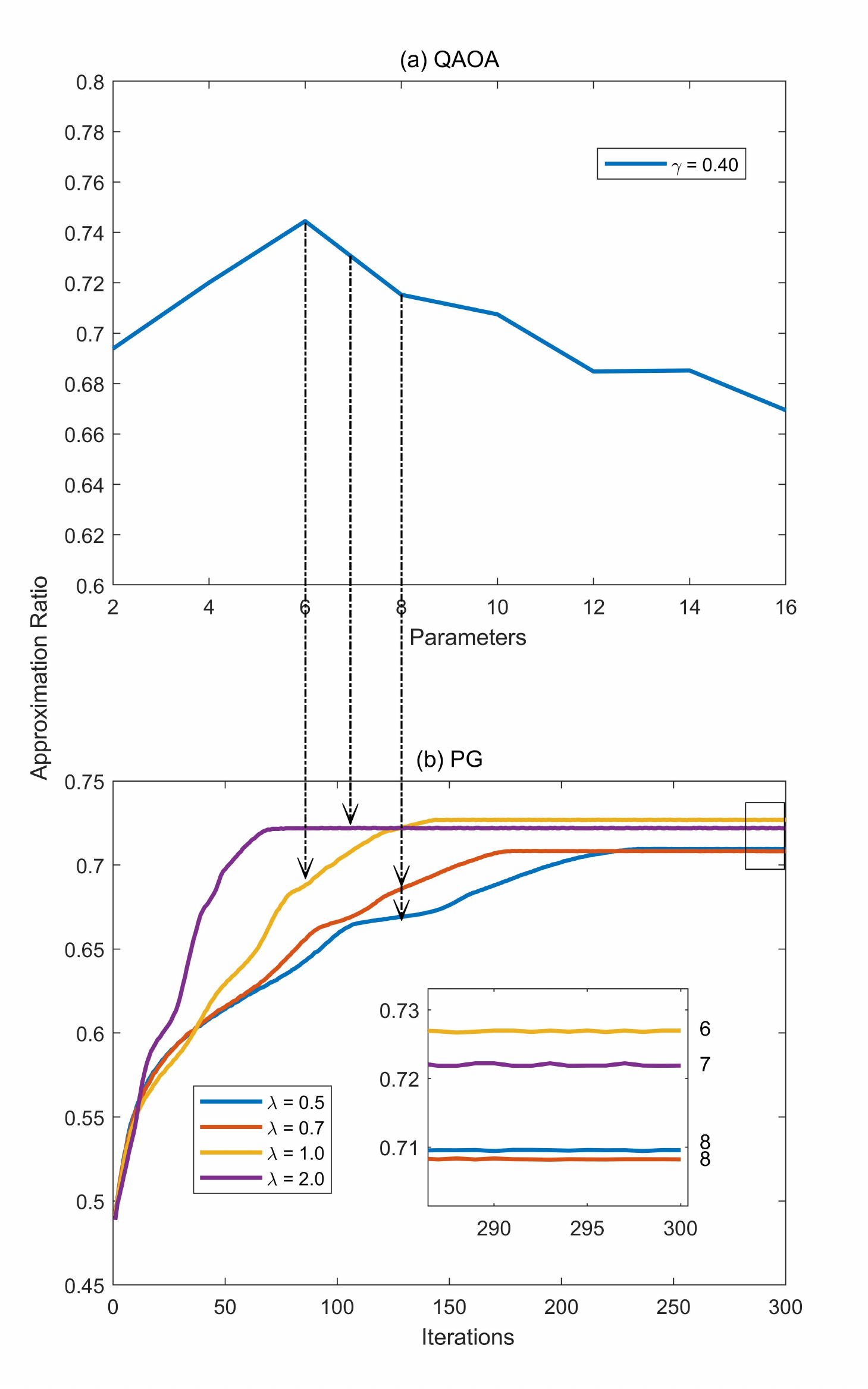}
	\caption{The performance comparison between (a) noisy QAOA and (b) PG algorithm under the influence of pure dephasing noise with the coupling strength being 0.4. Numbers on the right of the enlarged image in (b) are the finally selected parameters for different $\lambda$.}
	\label{fig.3}
\end{figure}

\subsection{Depth Selection with Dephasing Process}
The dephasing process is simulated by letting
\begin{equation}
L_1=...=L_n=\sigma_z.
\end{equation}

The experimental setup is the same as the relaxation process except that the coupling strength is set as 0.4. As shown in Fig.~\ref{fig.3}(a), the dephasing process has a strong influence on the approximation ratio, since the approximation ratio decreases rapidly after reaching the peak. In this case, the optimal depth is 6. As seen from Fig.~\ref{fig.3}(b), PG algorithm has successfully identified the optimal depth with $\lambda=1$. As the regularization strength decreases, the selected number of parameters increases in general, but the achieved approximation ratio will begin to decline if the reserved parameters are too many. These results demonstrate that depth selection algorithm maintains stable performance under various types of quantum noises.

\section{Conclusion}\label{Sec:C}
In this work, we have demonstrated that model selection algorithms such as PG can be applied to find the optimal depth of QAOA that yields the highest accuracy under realistic noises. These algorithms are efficient in the sense that only a few regularization parameters have to be tested before the optimal depth is found, which significantly reduces the number of experiments as compared to random search. In particular, it has been shown that the $l_1$-regularized algorithm is capable of determining the best depth with a sparsely distributed set of regularization parameters which will shrink by a constant factor after each round of experiment.

Although PG algorithm is specifically considered in this work, the proposed framework is compatible with any other model selection algorithms and regularization terms \cite{LUXBURG2011651}. Moreover, the optimization process does not rely on the information of the underlying dynamics of quantum systems, which makes it easy to implement on near-term quantum devices.

%\appendices
%% you can choose not to have a title for an appendix
%% if you want by leaving the argument blank
%\section{}
%% use section* for acknowledgment
%\section*{Acknowledgment}


%\ifCLASSOPTIONcaptionsoff
%  \newpage
%\fi

\section*{Acknowledgements}
This research was supported by the National Natural Science Foundation of China under Grant Nos. 62173296 and 61873162. G. Zhang also acknowledges support from Hong Kong Research Grant Council (Grants Nos. 1520841, 15203619, and 15506619), Shenzhen Fundamental Research Fund, China, under Grant No. JCYJ20190813165207290, and the CAS AMSS-polyU Joint Laboratory of Applied Mathematics.

\bibliographystyle{unsrtnat}
\bibliography{ref}

\end{document}